\magnification1200
\baselineskip14pt plus 2pt minus 1.1pt
\font\titlefont=cmb10 at 20pt
\font\authfont=cmr10 at 14pt
\font\affont=cmr10 at 12pt
\font\ninerm=cmr9
\font\nineit=cmti9
\font\eightrm=cmr8
\global\newcount\ftno \global\ftno=0

\catcode`\@=11
\def\f@@t{\baselineskip10pt plus 1pt minus 1pt \bgroup\ninerm\aftergroup\@foot\let\next}
\catcode`\@=12                                  
\def\foot{\global\advance\ftno by1\footnote{$^{\the\ftno}$}}

\centerline{\titlefont It was twenty years ago today $\ldots$}
\bigskip
\centerline{\authfont Paul Ginsparg}
\centerline{\affont Physics and Information Science, Cornell University}
\bigskip

\overfullrule0pt 
\noindent{\it To mark the 20$^{\rm th}$ anniversary of the commencement of hep-th@xxx.lanl.gov (now arXiv.org),
I've adapted this article from one\foot{Physics World, 1 Oct 2008; http://physicsworld.com/cws/article/print/35983\vskip-1pt
Learned Publishing, Vol.\ 22, No. 2, Apr 2009, p.\ 95; http://dx.doi.org/10.1087/2009203}
that first appeared in Physics World and was later reprinted (with permission) in Learned Publishing. 
This version is closer to my original draft, with some updates for this occasion, plus an astounding $2^5$ added footnotes.}\foot{Some updated information can also be found in P. Ginsparg, ``arXiv at 20'', Nature 476, 145--147 (11 Aug 2011); http://dx.doi.org//10.1038/476145a . More about the June 1991 Aspen activities, mentioned here on p.3, can be found in the slides of my June 2011 Aspen talk, entitled same as this, 
and available at http://people.ccmr.cornell.edu/\char'176ginsparg/blurb/asp11.pdf  .\vskip-11pt}

\bigskip

hep-th@xxx.lanl.gov received its first email submission on 14 Aug 1991.\foot{It was assigned the identifier 9108001,
now at http://arxiv.org/abs/hep-th/9108001}
Twenty years back is the timescale remembered (at least) as well as yesterday by those in mid-career, but viewed as ancient history by any generation of undergraduates.  And while each new generation thinks it's somehow unique, there are objective reasons to believe that the past two decades have witnessed an essential change in the way information is accessed, and how it is communicated to and from the general public, and among research professionals.

Mine was the first generation to have ready access to computers, starting in what was then known as junior high school in the late 1960s.
That meant a 100-baud teletype connected to a remote time-sharing system via an acoustically coupled modem, with paper punch tape as a storage medium for programs written in BASIC and PL/I.  By high school, I'd been exposed to Fortran programming on punch cards,\foot{at a summer program at Columbia University in 1971} submitted in batch mode for line-printer output the next day, and had the edifying experience of multiply reloading a boot sequence into a PDP-8's octal switches.\foot{at a summer program at Stonybrook University in 1972}
I first used email on the original ARPANET --- a predecessor of the Internet --- during my freshman year at Harvard University in 1973.\foot{to communicate with my brother, then a graduate student at Stanford}
My more business-minded classmates Bill Gates and Steve Ballmer were simultaneously strategizing ways to ensure that our class would have the largest average net worth of any undergraduate class ever.

Mine is also the last generation to have experienced the legacy print system, and I paid what was then known as a secretary\foot{Velma Ray, Hans Bethe's long-time assistant} to type my doctoral thesis at Cornell University in 1981.
The photocopy machine was a prime component of the distribution system back then, and I fondly recall teaching a recently retired Hans Bethe a thing or two about applied technology one slow weekend, by helping him to clear a paper jam.
But significant elements of change were already in the air in the late 1970s.
My thesis advisor Ken Wilson repeatedly promoted to us the need for massive parallel processing, and for the standardization of operating systems so that travelers to different institutions could immediately set to work without needing to learn a new interface.
In the early 1980s, he participated in the taskforce that advised the US National Science Foundation (NSF) to network together its soon-to-be-established supercomputer sites using the TCP/IP protocol.
That NSFNet backbone hastened the federation of existing networks, and sparked the dawn of the current Internet era.

Email usage became a more regular habit in the early 1980s, first within local computer systems and then via the growing primordial networks.
Back at Harvard in that period, I once explained with some effort to my colleague Sidney Coleman the then non-obvious phenomenon of receiving an email message via DECNet from the exterior, in this case from a former Harvard PhD student\foot{Orlando Alvarez} since moved to Berkeley.
Struggling to grasp the far-reaching implications, he furiously paced in a circle and, then, with dawning comprehension, presciently summarized\foot{to his then PhD student Phil Nelson and me}: ``The problem with the global village is all the global-village idiots.''

Following the appearance of Donald Knuth's TeXbook in 1984 --- the word-processing program that is still widely used to produce articles with mathematics content --- we switched en masse to computer typesetting our own articles.
The transition for the then-younger generation was virtually instantaneous, since the new methodology was an improvement in both process and quality of final result over what had preceded it, namely bribing a secretary to cut and paste with scissors and glue.
To facilitate cross-platform compatibility, Knuth intentionally chose plain text as TeX's underlying format, in addition providing a standard code for transmitting mathematical formulae in informal email communications.
Back and forth email exchanges would then frequently become the first draft of an article.
Nonetheless it took me real effort (and many years\foot{including a few years of intermittent connectivity ultimately  resolved by moving a ladder found leaning against a fibre optics cable in the basement of Cruft\vskip-11pt}) to get Harvard's physics department wired so that its VAX mainframe could be accessed from terminals in our offices.
The prevailing sentiment among the senior physics faculty was that their seminal work had been possible without computer access, and the desperate need of a digital crutch was no doubt evidence of the incorrigible feeblemindedness of a younger generation.

As the various pre-existing networks melded into the Internet by the late 1980s, email connectivity had reached critical mass in my own research community of high energy physics.
In those halcyon days, every message was from someone one knew personally, and contained useful content.
It was thus not common practice to advertise one's email address, but in late 1987 two collaborators and I first included our email addresses along with physical addresses in a preprint,\foot{L. Dixon, P. Ginsparg, J. Harvey, ``$\hat c=1$ Superconformal Field Theory'', scan available at
http://ccdb4fs.kek.jp/cgi-bin/img/allpdf?198808356} initiating that now-universal trend. When asked at that time, the dedicated librarians maintaining the essential SLAC-Spires bibliographic database\foot{specifically, Louise Addis} told me they would have loved to maintain on-line as well a full-text preprint database, but didn't have resources for the additional personnel required to solicit and handle electronic versions of articles; the now commonplace notion of automated repositories was still a few years in the future.

The exchange of completed manuscripts to personal contacts directly by email became more widespread, and ultimately led to distribution via larger email lists.\foot{The most significant of these was maintained by Joanne Cohn, then a postdoctoral associate at the IAS Princeton, who manually collected and redistributed preprints (originally in the subject area of matrix models of two dimensional surfaces) to what became a list of over a hundred
interested researchers, largely younger postdocs and grad students. This manual methodology provided an important proof of concept for the broader automated and archival system that succeeded it, and her distribution list was among those used to seed the initial hep-th userbase.\vskip-11pt}
The latter had the potential to correct a significant problem of unequal access in the existing paper-preprint distribution system.
For purely practical reasons, authors at the time used to mail photocopies of their newly minted articles to only a small number of people.
Those lower in the food chain relied on the beneficence of those on the A-list, and aspiring researchers at non-elite institutions were frequently out of the privileged loop entirely.
This was a problematic situation, because, in principle, researchers prefer that their progress depends on working harder or on having some key insight, rather than on privileged access to essential materials.

By the spring of 1991, I had moved to the Los Alamos National Laboratory, and for the first time had my own computer on my desk, a 25 MHz NeXTstation with a 105 Mb hard drive and 16 Mb of RAM.
I was thus fully cognizant of the available disk and CPU resources, both substantially larger than on a shared mainframe, where users were typically allocated as little as the equivalent of 0.5 Mb for personal use.\foot{i.e., 1000 blocks in some alternate system of units}
At the Aspen Center for Physics, in Colorado, in late June 1991, a stray comment from a physicist,\foot{Spenta Wadia}
concerned about emailed articles overrunning his disk allocation while traveling,\foot{The VAX operating system at the time had the inspired feature of counting one's incoming mail spool against one's disk allocation. One would typically delete just enough of the auxiliary files produced by TeX to be permitted to log off, then the next received email would result in being over quota, and subsequent emails would bounce without further alert. Most physicists are now either too young to know what a VAX was, or too old to remember.\vskip-11pt}
suggested to me the creation of a centralized automated repository and alerting system, which would send full texts only on demand.\foot{A month earlier, I had also read an article by another of my former (and future) Cornell mentors David Mermin
(``Publishing in Computopia'', Physics Today, May 1991, p.9), but twenty years later am unable to ascertain what subconscious effect it may or may not have had.  The letters in response (``The Rocky Road to Computopia", Physics Today, Jan 1992, p.13) were entertaining, disputing the possibility of a system by then already in existence. And his follow-up (``What's wrong in Computopia'', Physics Today, Apr 1992,  p.9) includes the memorable assertion that the hep-th system ``could well end up as [string theorists'] greatest contribution to science.''\vskip-11pt}
That solution would also democratize the exchange of information, leveling the aforementioned research playing field, both internally within institutions and globally for all with network access.

Thus was born xxx.lanl.gov,\foot{The name xxx was derived from the heuristic I'd used in marking text in TeX files for later correction (i.e., awaiting a final search for all appearances of the string `xxx', which wouldn't otherwise appear, and for which I later learned the string `tk' is employed by journalists, for similar reasons).  In those days of internet innocence, `xxx' had not yet acquired other connotations. Of course its now frequent appearance cost its former utility.\vskip-11pt} initially an automated email server  (and within a few months also an FTP server), powered by a set of csh scripts\foot{The csh scripts were translated to Perl starting in 1994, when NSF funding permitted actual employees.\vskip-11pt}
It was originally intended for about 100 submissions per year from a small subfield of high-energy particle physics,\foot{The name `hep-th' was suggested by Steve Shenker, based on recent experience establishing a ``String Institute'' at Rutgers\vskip-11pt} but rapidly grew in users and scope, receiving 400 submissions in its first half year.\foot{The short-term visibility and acceptance within the community were likely facilitated by the SLAC-Spires bibliographic database's use of the hep-th identifier scheme.  That resource, in turn, communicated its appreciation to have long-term persistent and consistent identifiers assigned to preprints, rather than just ephemeral institutional report numbers.\vskip-11pt}
The submissions were initially planned to be deleted after three months, by which time the pre-existing paper distribution system would catch up, but
by popular demand nothing was ever deleted.\foot{This is also the origin of the numbering scheme: after three months an automated process was to have executed `rm 9108$*$`, but known computerphobe physicist Andy Strominger argued within a month that the system was more convenient for archival retrieval than finding a photocopy in one's office, so the crontab was never installed.\vskip-11pt}
(Renamed in late 1998 to arXiv.org,\foot{The X was intended to evoke both the $\chi$ in $\TeX$ and `xxx'. My (future) wife suggested removing the final `e'.
The neologism was forced in part because Brewster Kahle had already registered archive.org and everything similar; but a decade later it proved quite useful, permitting, e.g., finding referrals to arXiv articles via news alerts with few false positives. (Even though, as I
later learned, it coincides with the Cyrillic spelling of the word for `archive' in Russian.)\vskip-11pt} it has accumulated roughly 700,000 total submissions [mid Aug 2011], currently receives  75,000 new submissions per year, and serves roughly one million full text downloads to about 400,000 distinct users per week. The system quickly attracted the attention of existing physics publishers, and in rapid succession I received congenial visits from the editorial directors\foot{Ben Bederson and Alan Singleton, respectively} of both the American Physical Society (APS) and  Institute of Physics Publishing (IOPP) to my little 10'x10' office.
It also had an immediate impact on physicists in less developed countries, who reported feeling finally in the loop, both for timely receipt of research ideas and for equitable reading of their own contributions.
(Twenty years later, I still receive messages reporting that the system  provides to them more assistance than any international organization.)

\smallskip
In the fall of 1992, a colleague\foot{Wolfgang Lerche} at CERN emailed me: `Q: do you know the worldwide-web program?' I did not, but quickly installed WorldWideWeb.app,  serendipitously written by Tim Berners-Lee for the same NeXT computer that I was using, and with whom I began to exchange emails.
Later that fall, I used it to help beta-test the first US Web server,\foot{Some of the early history of that resource is described here: H.B. O'Connell, ``Physicists
Thriving with Paperless Publishing'',  http://arxiv.org/abs/physics/0007040\vskip-11pt} set up by the library at the Stanford Linear Accelerator Center for use by the high-energy physics community.
Use of the Web grew quickly after the Mosaic browser was developed in the spring of 1993 by a group at the National Center for Supercomputer Applications at the University of Illinois (one of those supercomputer sites initiated a decade earlier, but poised to be replaced by massive parallelism), and it was not long before the Los Alamos `physics e-print archive' became a Web server as well.
Editorial control of the repository was barely necessary in those days, with the Internet still something of a private playground for academics, subject to few intrusions from the outside world.

Not everyone appreciated just how rapidly things were progressing.
In early 1994, I happened to serve on a committee advising the APS about putting Physical Review Letters online.
I suggested that a Web interface along the lines of the xxx.lanl.gov prototype might be a good way for the APS to disseminate its documents.
A response came back from another committee member: ``Installing and learning to use a WorldWideWeb browser is a complicated and difficult task ---
we can't possibly expect this of the average physicist.'' So the APS went with a different (and short-lived) platform. Meanwhile, the CERN website had partitioned its linked list of `all the web servers in the world' into geographic regions, as if keeping such lists could still be a sensible methodology for navigating information.

In the summer of 1994, Tim Berners-Lee, on his way out of CERN to found the World Wide Web Consortium at the Massachusetts Institute of Technology, kindly hosted me overnight at his home just over the French side of the border.
We discussed the implications of personal-computer chips suddenly leapfrogging heavy-duty workstations in performance, and the attendant dawning era of ubiquitous Web servers.
We marveled at how the Mosaic browser's support of inline graphics had transformed the perception of the Web's utility, and foreshadowed the rise of advertising.

During 1995, the penetration of our formerly private academic resources into the popular neocortex accelerated, with some form of `gee whiz' Internet news story almost every day: including how the WorldWideWeb had become the killer app, coupled with Netscape's public offering, the sky-is-the-limit futures of recent start-ups such as Yahoo, Time magazine's inevitable scare stories on the effects of cyberporn on children, and ending with 1995 being named the `year of the Internet' by Newsweek magazine.
While in Paris for a conference in 1996, I was struck by  all the the `http://$\ldots$' web URLs adorning the sides of vans and buses, signaling in a most public way the encroachment of commercial skyscrapers into our little academic playground.
The new `information superhighway' was heavily promoted for its likely impact on commerce and media, but the widespread adoption of social-networking sites facilitating file, photo, music and video sharing was not widely foreseen.

Fast-forwarding through the first dot-com boom and bust, and the emergent Google\-opoly, the effects of the technological transformation of scholarly communications infrastructure are now ubiquitous in the daily activities of typical researchers, lecturers, and students.
We have ready access to an increasing breadth of digital materials difficult to have imagined a decade ago.
These include freely available peer-reviewed articles from scholarly publishers, background and pedagogic material provided by its authors, slides used by authors to present the material, and videos of seminars or colloquia on the material --- not to mention related software, online animations illustrating relevant concepts, explanatory discussions on blog sites, often-useful notes posted by third-party lecturers of courses at other institutions, and collective wiki-exegesis.

A major lesson of the past decade has been that relatively simple algorithms and ample computing power applied to massive datasets result in resources whose utility far exceeds the naive sum of their conceptual components.
Web-search heuristics, hyperlinked journal references and citations, together  with search indexes,  the internet movie database,
social-networking sites, Amazon and other commercial sites, are all examples of this.
There are also threshold effects, in which seemingly minor improvements in software can have an overwhelming impact, for example using customized Web browsers\foot{i.e., moving from a stateless SMTP-based protocol on port 25 to a stateless HTTP-based protocol on port 80\vskip-11pt} instead of email transponders or FTP to access the same information repositories.
Similarly, blogs are fundamentally no different from the websites of a decade ago, but the pre-packaged software and tools for creating, linking, and maintaining them crossed some critical threshold and resulted in a new phenomenon. A few years ago,\foot{at SciFoo 2007, where the blogger was Aaron Swartz: http://www.aaronsw.com/weblog/scifoo07} glancing over my shoulder at a 20s-something blogging a seminar, I was struck by how a native laptop-user can navigate text and search windows faster than the eye can follow, and assemble references, photos, and graphics from multiple sources, simultaneously replying to comments, and in the end spend far less time to assemble a set of useful pedagogic pages, accessible to the entire world, than I spend writing problem-set solutions for a small class.
\xdef\scifoo{\the\ftno}

While looking to the future, it is also useful to assess some recent mistaken expectations.
In the mid-1990s, full-text searching appeared to many of us as a bootless exercise.
Search engines of the time --- such as AltaVista --- sort of worked due to the comparatively small amount of on-line information, but it was difficult to imagine that the methodology would scale as more information came on-line: if 10 times the number of pages meant that every query would bring up 10 times as many results, then any signal would be smothered by the overload.
But we've since learned that a relatively simple, yet nonetheless ingenious, set of heuristics can be used to order the search results, making use of the link structure of the Web\foot{i.e., PageRank, though the same methodology was employed earlier in a similar bibliographic context by G. Pinski and F. Narin,
``Citation Influence for Journal Aggregates of Scientific Publications: Theory, with Application to the Literature of Physics'',
Info.Proc.Man. Vol.\ 12, No.\ 5 (1976), pp.\ 297--312;  http://dx.doi.org/10.1016/0306-4573(76)90048-0 .
(By convention, things are named after the {\nineit last\/} person to discover them.) Others have pointed to similar methodologies employed in the slightly different context of econometrics models by economist W. Leontief in the 1940s (see M. Franceschet, ``PageRank: Standing on the shoulders of giants'';
http://arxiv.org/abs/1002.2858 ).\vskip-11pt} in addition to the text content of pages, so that for many typical queries the desired information appears among the top 10 results returned, and there is no need to peruse the many thousands of others.

That skeptical attitude regarding the potential efficacy of full-text searching carried over to my own website's treatment of crawlers as unwanted nuisances.
Seemingly out-of-control and anonymously-run crawls sometimes resulted in overly vociferous complaints to network administrators from the offending domain. A few years ago,$^{\scifoo}$ I was  reminded of a long-forgotten incident involving test crawls from some unmemorably named stanford.edu-hosted machines in mid-1996, when both Sergey Brin and Larry Page graciously went out of their way to apologize to me in person at Google headquarters for their deeds all those years ago.
Whatever was the memorable action taken by their system administrators, they were apparently not deterred for long. Ironically, looking back at the logs from that period uncovers their `problematic' traffic to have been entirely insignificant compared to that coming from any individual modern rss reader.

More recently, it was tempting to argue that a Wikipedia-like entity couldn't possibly work in the long run, that as soon as it became sufficiently popular it would devolve to a Usenet newsgroup cacophony of opinion and potential misinformation.
Yet after some publicly noted missteps, the primary Wikipedia site has evolved its policies to encourage academic practices such as citation of sources, and in the short-term remains surprisingly useful for a variety of academic and non-academic purposes.\foot{In addition to the lessons that web-scale search and crowdsourced resources could be made both feasible and useful, it was also perhaps unexpected
that users' social predilections would have them adopt en masse a mediocre user interface despite a privacy-violating financial model and lack of intrinsic hierarchical structure (e.g., Facebook), or that the nature and ubiquity of the mobile web would encourage so much communication in constrained 140 bite morsels (e.g., Twitter). A broader surprise is that so much of our electronic infrastructure is now funded by
advertising, adapting the model of commercial broadcast television from the previous century.\vskip-11pt}

In the direction of less-than-anticipated change, a decade and a half ago I certainly wouldn't have expected the current metastable state in physics publications, of preprint servers happily coexisting with conventional online publications, the two playing different roles. And it wasn't obvious two decades ago that a new generation of equation-intensive scholars would still be coding TeX by hand, without a proper WYSIWYG interface. In part, that is because newer methodologies have not been improvements in all relevant regards, as TeX was over its predecessors.

Physicists have been quick to adopt widespread pre-refereed distribution of scientific papers, but that has not been the case in other fields.
While quick and efficient information processing is a central component of scientific communication, scientific communities are also subject to internal social norms, which shape the use of new technologies. In the biomedical and life sciences, for example, adoption of preprint servers may be impeded by a long-standing tradition of regarding only refereed journal publication as a legitimate intellectual priority claim, together with concerns about public-health implications of the distribution of potentially misleading unrefereed results.

The new electronic infrastructure is moreover most frequently used as little more than a new means of distribution, and even the underlying document formats have not sufficiently evolved to take advantage of significant new opportunities.
We're only slowly moving from a situation in which the title, authors, references, and other dependencies of documents have to be guessed by cutting-edge artificial-intelligence techniques, to newer formats that automatically expose all such relevant metadata for standard query interfaces.  The current network benefits to readers will be increasingly shared by authors, as a new generation of network-aware authoring tools will analyze draft document content in progress, suggesting links to related external text and data resources, including semantic linkages.\foot{for a few more details and references, see 
P. Ginsparg, ``Text in a Data-centric World'', in ``{\nineit The Fourth Paradigm\/}'' (2009):\hfill\break
{\eightrm http://research.microsoft.com/en-us/collaboration/fourthparadigm/4th\_paradigm\_book\_part4\_ginsparg.pdf}\vskip-11pt}
 These will take advantage of the continued growth in distributed network databases, new interoperability protocols, machine-readable document standards, and relevant ontologies. Paraphrasing Marvin Minsky's once hypothetical backwards-looking comment regarding libraries and books, someone should soon ponder: ``Can you imagine they used to have an internet in which authors, databases, articles, and readers didn't talk to each other?''\foot{In {http://firstmonday.org/htbin/cgiwrap/bin/ojs/index.php/fm/article/view/864/773} ({\nineit The Battle to Define the Future of the Book in the Digital World\/}, First Monday, Vol.\ 6, No.\ 6, 4 Jun 2001), Clifford Lynch describes Minsky's original comment ``Can you imagine that they used to have libraries where the books didn't talk to each other?'' as ``simultaneously provocative, asinine, and inspiring.''\vskip-11pt}

Scholarly journals were the earliest instantiation of `Web 2.0' methodology, insofar as it describes the deployment of some skeletal infrastructure into which users deposit content, the value of which in turn is increased by general accessibility.
But the scholarly community has been slower to incorporate the latest round of social-networking tools into its regular practices.
When the Internet was essentially an academic monopoly, new developments were naturally adapted to the needs of the research community.
The focus is now elsewhere, and the vast resources invested in commerce and entertainment have left scientists momentarily behind the forefront of interactive Web phenomena.
The very nature of scholarly pursuits leaves academics slightly displaced from the bleeding 
edge, with the shift of the centre of mass towards popular consumption resulting in an ever-smaller percentage of new resources directed, or well adapted, to those pursuits.

Many useful lessons can nonetheless be inferred from the popular arena.
For example, no legislation is required to encourage users to post videos to YouTube, whose incentive of instant gratification, through making personal content publicly available, parallels the scholarly benefit of voluntary participation in the incipient version of arXiv.org in 1991.
If scholarly infrastructure can be upgraded to encourage maximal spontaneous participation, then we can expect not only increasingly automated interoperability among databases and increasing availability of materials online for algorithmic harvesting --- articles, datasets, lecture notes, multimedia, and software --- but also qualitatively new forms of academic effort.
Expertise-intensive tags, links, comments, corrections, contributions to ontologies, and linkages, all actively curated, will become increasingly important, acting to glue databases and texts together into a more powerful knowledge structure.
Such work will need to be credited as scholarly achievement, along with the future analog of conventional journal publication.

Scholarly infrastructure will employ as well a passive ingest of readership, bookmarking and annotation behavior, meshed together with the above active component in a more bottom-up approach to quality control. 
The goal is the creation of a semi-supervised and self-maintaining knowledge structure, navigated via synthesized concepts, cleaned of redundancy and ambiguity, sourced, authenticated, and highlighted for novelty.
Our browsing of the literature will be far more comprehensive, guided by algorithms with access to our own and collective user behaviors; and our reading of individual components that much more incisive, guided by linkages to explanatory and complementary resources tied to words, equations, figures, and data.

The result will be a transformation in the way we process scientific information, much as the availability of interlinked network resources has led to new non-linear reading strategies, and the availability of networked mobile devices has altered the way we use our short- and long-term memories.
The Internet, World Wide Web, search engines, and other developments described here all initially stemmed from the academic community's need to transmit, retrieve, and organize information.
It is exciting to project that new research and cognitive methodologies to be developed for academic use may ultimately be adopted as well by the general public for the creation and dissemination of knowledge.

\bye